# Uncertainty relations for incompatible observables: Newer results versus modified strategies

Kamal Bhattacharyya[*]

Department of Chemistry, University of Calcutta, Kolkata 700 009, India

**Abstract**

Two special situations where the standard uncertainty product inequality appears to be useless are modified. One such case is noted to also trivialize the recently-introduced alternatives [Phys. Rev. Lett. **113**, 260401 (2014); Sci. Rep. **6**, 23201 (2016)] involving sums of variances. A careful discussion is presented on the experimental justifications of some of the relations [Phys. Rev. A **93**, 052108 (2016)] using qutrit and qubit states. Alternative bypass routes are put forward to tackle this situation, with and without involving any auxiliary state. This latter strategy is noted to be vital in an entirely different context concerned with the quality of approximate stationary states. The other case is more frustrating, but an effective method is advanced. En route, the recent alternatives are also simplified to easily accommodate even the cases of more than two observables. In favorable circumstances, an easy option in function space is obtained by virtue of symmetry that does not involve any auxiliary state. Pilot calculations reveal the advantages of our endeavor.

**Keywords:** Uncertainty product; Sums of variances; Qutrit; Qubit

**PACS numbers:** 03.65.Ta, 42.50.Lc

# I. INTRODUCTION

Given two incompatible observables $A$ and $B$ (*i.e.*, $[A, B] \neq 0$) and a normalized state $\Psi$, one can define

$$\Psi_A = (A - \langle A \rangle I)\Psi, \langle A \rangle = \langle \Psi | A | \Psi \rangle, \langle \Psi_A | \Psi_A \rangle = (\Delta A)^2. \quad (1)$$

Definition (1) allows us to identify another normalized state

$$\bar{\Psi}_A = \Psi_A / \Delta A$$

that satisfies $\langle \Psi | \bar{\Psi}_A \rangle = 0$. A similar set of relations holds for observable $B$. Applying the standard Cauchy-Schwarz inequality to states $\Psi_A$ and $\Psi_B$, one obtains

$$\langle \Psi_A | \Psi_A \rangle \langle \Psi_B | \Psi_B \rangle = (\Delta A)^2 (\Delta B)^2 \geq |\langle \Psi_A | \Psi_B \rangle|^2. \quad (2)$$

On expansion of the right hand side of (2), the oft quoted form (3) of the uncertainty product inequality [1-4]

$$(\Delta A)^2 (\Delta B)^2 \geq \left|\tfrac{1}{2}\langle [A, B]\rangle\right|^2 + \left(\tfrac{1}{2}\langle [A, B]_+\rangle - \langle A\rangle\langle B\rangle\right)^2 \quad (3)$$

follows. This textbook [5] version has continued to encourage refinements along several directions. Non-hermitian operators were included [6-7] in specific situations, 'local' uncertainties were constructed [8] to get added insights, the connection with quantum information was established [9] and pathological cases were carefully analyzed [10-11], among others. The effects of measurement error, noise and disturbance have also been reviewed in detail [12-13]. Other works have focused attention on non-locality [14], and extension of the variance-based uncertainty relation to include more than two observables [15-17]. Noting that the variance is not always a good measure of the 'width' of a distribution [9, 18-19], a different class of works highlighted the entropic formulation [20 - 24] as a potential alternative. Perspectives of this form are available [19, 25], along with historical notes and application to cryptography [25]. A recent reformulation [26] has discussed both the approaches.

Inequality (2) is saturated only when $\Psi_A = \mu \Psi_B$, where $\mu$ is a *c*-number Thus, it nicely distinguishes linearly independent $\Psi_A$ and $\Psi_B$ from the linearly dependent ones. Indeed, one can separate three clear cases in respect of (2):

**Case 1:** $\Psi_A = \mu \Psi_B$, or $\langle \Psi_A | \Psi_B \rangle \neq 0$ .

**Case 2:** $\Psi_A = 0$ or $\Psi_B = 0$.

**Case 3:** $\langle \Psi_A | \Psi_B \rangle = 0$, but $\Psi_A \neq 0, \Psi_B \neq 0$.

Of these, the last two situations are problematic in relation to the standard uncertainty product inequality.

To specifically tackle Case 2, considerable recent interest [27-29] has been paid to 'stronger' uncertainty relations in the form of *sums of variances*. Thus, a few new inequalities [27, 28] have appeared [see Eqs. (4) – (5) below]. The nature and spirit of these relations are, however, very different from (2) because they involve *auxiliary states* that can be varied to improve the bounds. Experimental relevance of some of these inequalities involving qutrit/qubit states [29] has also emerged. They have received immediate attention too [16-17, 26].

In the present communication, we intend to first explore the fate of these sum-form inequalities in Case 2 [Sec. IIB1, Eq. (6)]. Some remarks on the experimental realizations are also made [Sec. IIB2]. Secondly, we provide two routes in Sec. IIB3 and Sec. IIB4 to overcome the problem, the latter without involving any auxiliary state, and hence more appealing. Its impact in the context of estimating the quality of approximate stationary states is also briefly outlined. Thirdly, we concentrate on Case 3 and advance a general scheme to overcome the problem associated with (2). Further analysis leads to a few simpler forms, displayed in Eqs. (14) – (17). We may employ any of these relations to handle Case 3. Particularly, we advocate our product form with just one auxiliary state [Eq. (17)] that is found to be the simplest, effective and easily extendable to more than two observables as well. Finally, a special situation is highlighted where symmetry of the potential simplifies the issue in Case 3, requiring no auxiliary state.

## II. ANALYSIS OF THE VARIOUS CASES

### A. Case 1

Simplest is to deal with this case, and it is most common too. The inequality for linearly independent $\Psi_A$ and $\Psi_B$ reduces to equality for linearly dependent states. Note, however, that (i) Eq. (2) does not require any auxiliary state anywhere, (ii) it yields a finite answer for the right side, and (iii) in case the right side of (2) becomes zero, each term at the right of (3) will also be zero.

### B. Case 2

If $\Psi_B = 0$, the way (2) is derived reveals *only* that $\Delta A \geq 0$. Yet, at $\Delta B = 0$, there should really be no problem with (2) as such, because the term 'uncertainty product' does not carry any meaning with *zero* spread for observable $B$; one is free to calculate $\Delta A$ independently, if the quantity exists [10, 11]. However, the new inequalities lay emphasis on the 'sum' form. Indeed, kinship of such a form with (2) has long been known (see, *e.g.*, [5], problem 9-6). Here, the forms at issue are [27]

$$(\Delta A)^2 + (\Delta B)^2 \geq \pm i \langle [A, B] \rangle + \left| \langle \Psi_A | \Psi^\perp \rangle \pm i \langle \Psi_B | \Psi^\perp \rangle \right|^2 \tag{4a}$$

$$(\Delta A)^2 + (\Delta B)^2 \geq \tfrac{1}{2} \left| \langle \Psi_A | \Psi^\perp_{A+B} \rangle + \langle \Psi_B | \Psi^\perp_{A+B} \rangle \right|^2. \tag{4b}$$

In (4a), the overall sign of the first right hand term should be chosen as positive. The normalized states $\Psi^\perp$ satisfy $\langle \Psi | \Psi^\perp \rangle = 0$. The state $\Psi^\perp_{A+B}$ in (4b) would not only be orthogonal to $\Psi$, but proportional to $(\Psi_A + \Psi_B)$ also. Two other inequalities, respectively more general than (4a) and (4b), have appeared very recently [28], *viz.*,

$$(1+\lambda)(\Delta A)^2 + (1+\lambda^{-1})(\Delta B)^2 \geq \pm 2i \langle [A, B] \rangle + \left| \langle \Psi_A | \Psi^\perp_1 \rangle - i \langle \Psi_B | \Psi^\perp_1 \rangle \right|^2$$
$$+ \lambda^{-1} \left| \lambda \langle \Psi_A | \Psi^\perp_2 \rangle - i \langle \Psi_B | \Psi^\perp_2 \rangle \right|^2, \tag{5a}$$

$$(1+\lambda)(\Delta A)^2 + (1+\lambda^{-1})(\Delta B)^2 \geq \left| \langle \Psi_A | \Psi^\perp_{A+B} \rangle + \langle \Psi_B | \Psi^\perp_{A+B} \rangle \right|^2$$
$$+ \lambda^{-1} \left| \lambda \langle \Psi_A | \Psi^\perp \rangle - \langle \Psi_B | \Psi^\perp \rangle \right|^2. \tag{5b}$$

Both the results in (5) are true for a free parameter $\lambda$ that can be varied to gain extra advantage (see, e.g., [28]), with already defined states, and states satisfying $\langle \Psi | \Psi^\perp_j \rangle = 0$ for $j = 1, 2$. However, these relations refer to *weighted uncertainties* (for $\lambda \neq 1$) that are not of usual interest.

### *1. Simplifications*

A few quick observations may now be in order: (i) If $\Psi_B = 0$, and hence $\Delta B = 0$, say, which appeared as one of the *most interesting* situations [27 - 29], all of (4a), (5a) and 5(b) reduce to the form

$$\Delta A \geq \left| \langle \Psi_A | \Psi^\perp \rangle \right|, \tag{6}$$

while 4(b) is weaker by a factor of $1/\sqrt{2}$ in (6). But, relation (6) is practically obvious because of the trail

$$\langle \overline{\Psi}_A | \overline{\Psi}_A \rangle = 1 \Rightarrow \langle \overline{\Psi}_A | \Psi_N \rangle \leq 1 \Rightarrow |\langle \Psi_A | \Psi_N \rangle| \leq \Delta A. \qquad (7)$$

(ii) In (7), we have chosen $\Psi_N$ as some arbitrary normalized state; any component of $\Psi$ present in $\Psi_N$ will automatically be projected out by $\Psi_A$. Hence, the equivalence of (6) and (7) follows; equality is achieved in either situation by choosing $\Psi^\perp = \Psi_N = \overline{\Psi}_A$. (iii) Relations (4) and (5), unlike (2), have a variational character. Hence, they *do not* distinguish situations where $\Psi_A$ and $\Psi_B$ are linearly dependent from the linearly independent ones. To state otherwise, *equality* in (4) or (5) is not automatically ensured for $\Psi_A = \mu \Psi_B$. A beautiful feature of (2) is thus lost. (iv) Most significantly, (2) works with *known* states, but the alternative forms (4) and (5) require *at least one auxiliary state*. Indeed, if we exploit the *full* freedom of the auxiliary state in (4a), the $\lambda$-dependent relations in (5) become unimportant.

## 2. The experimental relevance

We now scrutinize the experiment [29] with qutrits to justify (4a) and (4b). First, they trivialized (2) by choosing either $\Psi_A = 0$ ($A = J_x$) or $\Psi_B = 0$ ($B = J_y$) at some point. Explicitly, they focused attention on the state

$$\Psi = (\sin\varphi \quad 0 \quad \cos\varphi)^T, \varphi \in [0,\pi]. \qquad (8)$$

at $j = 1$. The two other states, and related quantities of interest, (with standard [5] representations for the operators and $\hbar = 1$) are as follows:

$$\Psi_A = \tfrac{1}{\sqrt{2}}(\sin\varphi + \cos\varphi)(0 \quad 1 \quad 0)^T ; \Psi_B = \tfrac{-i}{\sqrt{2}}(\cos\varphi - \sin\varphi)(0 \quad 1 \quad 0)^T ;$$
$$\langle J^2 \rangle = 2, \langle J_z^2 \rangle = 1, \langle J_x \rangle = \langle J_y \rangle = 0. \qquad (9)$$

Thus, when $\varphi$ is chosen as (say) $\pi/4$, $\Psi_B = 0$. So, secondly, they obtained at this point only partial results from (4a) and (4b), viz. $(\Delta J_x)^2 \geq (\Delta J_x)^2$ and $(\Delta J_x)^2 \geq (\Delta J_x)^2 / 2$, respectively, in conformity with what we mentioned around (6). Third, their verification of the equality in 4(a), viz., $(\Delta J_x)^2 + (\Delta J_y)^2 = 1$ at *any* $\varphi$ (see figure 2 of ref. [29]), is obvious here because of results (9) for the chosen state $\Psi$ in (8), coupled with the standard relation $J_x^2 + J_y^2 = J^2 - J_z^2$. In effect, the exact answer is known beforehand without calculating any variance. In Appendix A, we summarize the specialties of this chosen state with reasons. Fourth, choice of (8) is quite pathological *at $\varphi = \pi/4$*. At such a point, Ψ becomes an eigenstate of all the three operators $J^2$, $J_z^2$ and $J_y$, respectively with eigenvalues 2, 1 and 0 [see Appendix A, around Eqs. (A11) and (A12)]. Again, such an abnormal situation crops up at $\varphi = 3\pi/4$. Fifth, $\Psi_A$ and $\Psi_B$ are linearly dependent in general.

The experiment [29] on qubits with $A = \sigma_x$ and $B = \sigma_y$ revealed again virtually similar features. The chosen state and other relevant quantities are displayed below:

$$\Psi = (\sin\varphi \quad \cos\varphi)^T; \Psi_A = \cos 2\varphi (\cos\varphi \quad -\sin\varphi)^T; \Psi_B = -i(\cos\varphi \quad -\sin\varphi)^T;$$
$$\langle\sigma_x\rangle = \sin 2\varphi, \langle\sigma_y\rangle = 0; \Delta\sigma_x^2 + \Delta\sigma_y^2 = 1 + \cos^2 2\varphi. \quad (10)$$

It shows, *at* $\varphi = \pi/4$, $\Psi_A = 0$, and hence $\Delta A = 0$. Therefore, almost all our previous observations would apply (note, however, missing of a scale factor of 2 in the ordinate of figure 4 in ref. 29; see also Appendix A). Like the qutrit case, an additional peculiarity shows up at $\varphi = 3\pi/4$. In view of the set of known relations, $\sigma_x^2 = \sigma_y^2 = \sigma_z^2 = I$, here too, one need not calculate any squared average. Note that $\Psi_A$ and $\Psi_B$ are generally linearly dependent again.

With two linearly dependent states ($\Psi_A$ and $\Psi_B$), we know that inequality (2) is *saturated*. For the above two examples, one obtains neatly similar answers:

$$\begin{aligned}\Delta J_x \Delta J_y &= \tfrac{1}{2}|\cos 2\varphi| \\ \Delta\sigma_x \Delta\sigma_y &= |\cos 2\varphi|\end{aligned}. \quad (11)$$

Being an *equality*, the right side of (2) thus offers the *best* possible result in either situation at *any* $\varphi$, barring the two special points.

### *3. The first resolution*

In view of our simplification, a straightforward resolution via (7) would be to employ $\Delta A \geq |\langle\Psi_A | \Psi_N\rangle|$. Any normalized $\Psi_N$ can work here. The inequality is saturated for the choice $\Psi_N = \overline{\Psi}_A$. However, there should be absolutely no problem in directly estimating $\Delta A$ here because both the state and the operator in hand are known.

### *4. A second route*

If $[A, B] = iC$ and $\Delta B = 0$, i.e., $B\Psi = \beta\Psi$, a more interesting alternative is to go for the operator $\overline{B} = B - A$, in conjunction with $A$. While we still have $[A, \overline{B}] = iC$, the redefined operator has the advantage of revealing that $\Delta A = \Delta\overline{B} \neq 0$. Hence, the left side of (2) will read as $\Delta A \Delta \bar{B} = (\Delta A)^2$, and the right side will *not* vanish. Here, $\Delta A$ can be directly estimated, and this is a better option; else, a bound to it is obtainable via our prescription in Sec. IIB3, outlined just above. Origin of the equality referred to above lies in the relation $\Psi_{B-A} = -\Psi_A$. Notably, this prescription *avoids* also the use of *any* auxiliary state. A different kind of possible alternative might have been the relation

$$\Delta A + \Delta B \geq \max\{\Delta(A+B), \Delta(A-B)\},$$

as given by Eq. (7) in Ref. 16. While such an inequality is devoid of any auxiliary state, it too is useless when $\Delta B = 0$. Hence, we stick to the use of $\bar{B} = B - A$ to proceed forward.

Let us now apply the above bypass route to the observable $J_y \equiv B$, as mentioned, and consider the qutrit case discussed in Sec. IIB2. When $\varphi = \pi/4$ in (8), we shall accordingly choose

$$\bar{B} = J_y - J_x = -\tfrac{1}{2}[J_+(1+i) + J_-(1-i)]$$

to obtain $\Delta \bar{B} = \Delta J_x$, and thus the problem disappears. At the other point, $\varphi = 3\pi/4$, one would simply exchange the roles of *A* and *B*. The trouble with qubits for the state in (10) at the two specific points may likewise be overcome.

Another very common place consequence of Case 2 is that, any eigenstate of the Hamiltonian *H*, defined by $H\Psi = (T+V)\Psi = E\Psi$, obeys $\Delta V \Delta H = 0$ and $\Delta T \Delta H = 0$, though [*V*, *H*] ≠ 0 and [*T*, *H*] ≠ 0. However, use of the redefined operator like $\bar{B} = (H-T)$ immediately reveals via our route that the state has to satisfy $\Delta T = \Delta V$, *i.e.*, the *kinetic and potential energy uncertainties are equal for any stationary state*. This relation is *independent* of state (quantum number) and possesses some sort of a universal character. A physical reason is that, since the total energy is precise (*B* ≡ *H*), any deviation from the average value in the kinetic energy part would be exactly counterbalanced by the same in the potential energy part.

Incidentally, the departure from such a relation like $\Delta T = \Delta V$ between the standard deviations may also be fruitfully employed to assess the quality of *approximate stationary states*. The role of electron correlation [30] is a case in point. One may note, this measure is far more stringent than the virial theorem [31], yet simpler than either the Eckart criterion [32] or the method of local energy [33]. Whereas the latter approach often turns out to be messy, finding error bounds to the average energy is no less challenging [34]. Quite a few of such techniques have been discussed elsewhere [35] with pertinent references. However, it is easy to check here the efficacy of the present measure even by considering a very simple problem [36] where a chosen variational state is *optimally scaled* so that the virial theorem is *exactly* satisfied [37], though the state is *not* truly *stationary*. This becomes immediately apparent when we compare the estimates of $\Delta T$ and $\Delta V$ [36].

### C. Case 3

The real challenge to (2) is *not* posed by $\Psi_A = 0$ or $\Psi_B = 0$; rather, it refers to cases where we have a different restriction like

$$\langle \Psi_A | \Psi_B \rangle = 0, \qquad (12)$$

so that it [or (3)] will fail to yield a lower bound better than zero (though the equality sign does not apply). Notice that definition (1) ensures $\langle \Psi_A | \Psi \rangle = 0, \langle \Psi_B | \Psi \rangle = 0,$ where $\Psi$ is the given, but otherwise arbitrary, state. Therefore, at the onset, three mutually orthogonal states are identified. The task is to obtain a *non-zero* lower bound for (2).

#### 1. A general resolution using auxiliary functions

Here, we need to *import* auxiliary state(s). First, let us look back at (7). For two observables, it can be cast as

$$\Delta A \Delta B \geq |\langle \Psi_A | \Psi_N \rangle| |\langle \Psi_B | \Psi_N \rangle| \qquad (13)$$

where the equality will hold either for $\Psi_A = 0$ or $\Psi_B = 0$, or when $\Psi_B = \mu \Psi_A$ and $\Psi_N = \bar{\Psi}_A$. The corresponding sum form will then read much simpler than (4) or (5):

$$\Delta A + \Delta B \geq |\langle \Psi_A | \Psi_N \rangle| + |\langle \Psi_B | \Psi_N \rangle|. \qquad (14)$$

If, on the other hand, $\Psi_A$ and $\Psi_B$ are linearly independent or orthogonal, (13) or (14) may not offer sufficiently tight bounds. One may go for inequalities involving *two auxiliary states* as in (5), *e.g.*, of the form

$$\Delta A \, \Delta B \geq |\langle \Psi_A | \Psi_{N1} \rangle| |\langle \Psi_B | \Psi_{N2} \rangle| \qquad (15)$$

that modifies (13), where $\Psi_{N1}$ and $\Psi_{N2}$ are arbitrary normalized states. In the other case (14), it generalizes to

$$\Delta A + \Delta B \geq |\langle \Psi_A | \Psi_{N1} \rangle| + |\langle \Psi_B | \Psi_{N2} \rangle|. \qquad (16)$$

Thus, (14) and (16) come up as *simpler* alternatives to (4) or (5). Advantages of incorporating variational parameters in $\Psi_N, \Psi_{N1}, \Psi_{N2}$ are obvious. Needless to mention, (16) is saturated for the choice $\Psi_{N1} = \bar{\Psi}_A$ and $\Psi_{N2} = \bar{\Psi}_B$ (see Appendix B). Additionally, forms (13) and (15) *modify* the time-honored form (2). We record also in passing that, when (12) holds, (13) may be strengthened further to yield (see Appendix B)

$$\Delta A \Delta B \geq 2 |\langle \Psi_A | \Psi_N \rangle| |\langle \Psi_B | \Psi_N \rangle|. \tag{17}$$

Since (17) necessitates just *one* auxiliary state, and only (12) trivializes (2), we shall henceforth be chiefly concerned with it. This is the key relation *that extends the applicability* of (2). It is saturated only if $\Psi_N = \frac{1}{\sqrt{2}}(\bar{\Psi}_A + \bar{\Psi}_B)$. Further, inequality (15) or (17) is extendable to more than two incompatible observables [15 – 17] as well. For example, form (17) can be generalized (see Appendix B) to read as

$$\Delta A \Delta B \, \Delta C ... \geq n^{n/2} |\langle \Psi_A | \Psi_N \rangle| |\langle \Psi_B | \Psi_N \rangle| |\langle \Psi_C | \Psi_N \rangle|...$$

for *n* mutually orthogonal states $\Psi_A$, $\Psi_B$, $\Psi_C$, etc. Such an advantage of extension using merely one auxiliary state is lacking in the sum form (16).

## *2. Examples*

Let us cite two simple examples to justify the endeavor. The qutrit case will be chosen first, since considerable attention to such states has already [27 - 29] been paid. Qubits do not enter the present discussion because, being a 2x2 problem, one cannot have three orthogonal states here. Finite-dimensional matrix eigenvalue equations do not qualify as very general quantum-mechanical problems too. So, our other situation is picked up from proper quantum mechanics. It also deserves special attention due to a unique possible bypass route that *does not* require *any auxiliary* state.

**Example 1:** Consider again the operators $J_x$, $J_y$ and $J_z$, with $J_x$ as *A* and $J_y$ as *B*. The state is taken now, in place of (8), as

$$\Psi = \tfrac{1}{\sqrt{3}} \begin{pmatrix} 1 & 1 & 1 \end{pmatrix}^T \tag{18}$$

to obtain the following results:

$$\langle J_x \rangle = 2\sqrt{2}/3, \langle J_y \rangle = 0, \langle J_z \rangle = 0, \langle J_x^2 \rangle = 1, \langle J_y^2 \rangle = 1/3;$$
$$\Psi_A = \tfrac{1}{3\sqrt{6}} \begin{pmatrix} -1 & 2 & -1 \end{pmatrix}^T ; \Psi_B = \tfrac{1}{i\sqrt{6}} \begin{pmatrix} 1 & 0 & -1 \end{pmatrix}^T. \tag{19}$$

Clearly, (2) is trivialized here because of (12). So, we have to avoid the disaster. At this juncture, we also notice that any such state satisfies

$$\langle \Psi | [A, B] | \Psi \rangle = 0$$

and hence the first factors at the right sides of both (4a) and (5a) will vanish. Anyway, to proceed, we choose two *arbitrary* normalized states as

$$\Psi_{N1} = \tfrac{1}{\sqrt{2}}(-1 \quad 1 \quad 0)^T, \Psi_{N2} = \tfrac{1}{\sqrt{3}}(1 \quad -1 \quad -1)^T. \tag{20}$$

For a thorough comparison, let us take (4a), (4b), (5a), (5b), (13), (14), (15), (16) and (17), and display their right side performances in Table 1. For convenience, values of the relevant integrals are summarized below:

$$\langle \Psi_A | \Psi_{N1} \rangle = 1/2\sqrt{3}; \langle \Psi_B | \Psi_{N1} \rangle = i/2\sqrt{3}; \langle \Psi_A | \Psi_{N2} \rangle = -\sqrt{2}/9; \langle \Psi_B | \Psi_{N1} \rangle = -i\sqrt{2}/9.$$

**Table 1. A comparative performance test of some inequalities: Estimates from the right hand sides (RHS) of selected equations from the text are provided. Use of $\Psi_{N1} = \Psi_N$ is denoted by (i) and $\Psi_{N2} = \Psi_N$ is denoted by (ii), along with equation numbers in the table, wherever appropriate. For (5a) and (5b), $\Psi_{N1}$ is taken as the first $\Psi_N \equiv \Psi^\perp$ state and $\Psi_{N2}$ as the second such state.**

| Eq. | RHS | Eq. | RHS |
| --- | --- | --- | --- |
| 4a (i) | 1/3 | 13 (ii) | 2/27 |
| 4a (ii) | 32/81 | 14 (i) | 1/√3 |
| 4b (i) | 1/12 | 14 (ii) | 4√2/9 |
| 4b (ii) | 10/81 | 15 | 1/(3√6) |
| 5a | 59/81 | 16 | (√3+2√2)/6 |
| 5b | 67/162 | 17 (i) | 1/6 |
| 13 (i) | 1/12 | 17 (ii) | 4/27 |

Our comparative survey includes (5a) and (5b), but at $\lambda = 1$, so that no extra weight is put on any dispersion and these inequalities can compete with others. Let us remark here that, as prescribed [28], (5a) does not reduce to (4a) at $\lambda = 1$, *unless* we *also* demand that $\Psi^\perp_{N2} = \Psi^\perp_{N1} = \Psi^\perp_N$, losing deliberately an added flexibility. This continues to be true for (5b) as well. However, results will differ at $\lambda = 1$ only when we choose two different states in (5). In this way, estimates unlike (4a) and (4b) are found, and they are displayed in Table 1. An initial comparison reveals that (*a*) Eq. (17) (i) performs best among all the product inequality forms (13), (15) and (17). (*b*) The performances of (15) and (16) are better than (13) and (14), respectively, as already remarked [see below (14)]. Thus, in our sum form containing standard deviations, (16) excels. (*c*) If we now compare the results among (4a), (4b), (5a) and (5b) involving sums of variances, we note that (4a) (ii) stands out. Thus, the following three inequalities are worth mentioning in the present context:

$$\begin{aligned} &(17)\ (i): \Delta J_x \Delta J_y \geq \tfrac{1}{6}; \\ &(4a)\ (ii): \Delta J_x^2 + \Delta J_y^2 \geq \tfrac{32}{81}; \\ &(16): \Delta J_x + \Delta J_y \geq \tfrac{\sqrt{3}+2\sqrt{2}}{6}. \end{aligned} \qquad (21)$$

It may be interesting to compare the direct estimate from (16) [the third entry in (21)] with a value derived from the other two best values mentioned above it. This is obtained as follows:

$$\Delta J_x + \Delta J_y \geq \sqrt{\tfrac{32}{81} + 2\tfrac{1}{6}}.$$

But, we clearly see that even this bound is inferior to the result found from (16). In passing, therefore, we happily note that both (16) and (17) clearly win to show the benefit of simplicity. Moreover, while (16) establishes itself as a better alternative to (4) or (5), (17) bypasses (2) quite fruitfully when (12) holds. Indeed, the latter is noted to perform best in the issue at hand. At the same time, close performance of some other recipes put forward by us may also be found from Table 1. For example, (17) (ii) is marginally inferior to (17) (i), and this is because of the slightly better overlap of $\Psi_{N1}$ than $\Psi_{N2}$ with the particular $\Psi_N$ that leads to (B5) [see Appendix B].

**Example 2:** We next take up the case where $A = x^2$ and $B = p$. Our system is the one-dimensional simple [5] harmonic oscillator described by the Hamiltonian

$$H = p^2 + x^2,\ \hbar = 1, m = \tfrac{1}{2}, K = 2, \qquad (22)$$

where $K$ is the force constant. $H$ in (22) satisfies $H\phi_j = (2j+1)\phi_j$; $\langle \phi_j | \phi_j \rangle = 1$. The state $\Psi$ is chosen as $\phi_0$ to find

$$\begin{aligned} &\langle x^2 \rangle = 1/2, \langle p \rangle = 0, \langle x^4 \rangle = 3/4, \langle p^2 \rangle = 1/2; \\ &\Psi_A = \left(x^2 - \langle x^2 \rangle\right)\phi_0 = \tfrac{1}{\sqrt{2}}\phi_2;\ \Psi_B = \left(p - \langle p \rangle\right)\phi_0 = \tfrac{i}{\sqrt{2}}\phi_1. \end{aligned} \qquad (23)$$

Note that, two other distinctly different normalized eigenstates of $H$ emerge. So, (12) is satisfied. Therefore, the need to use (17) is obvious. To achieve this end, here we take

$$\Psi_N = \tfrac{1}{\sqrt{(1+\eta^2)\pi}}\left(\sin x + \eta \cos \tfrac{3}{2}x\right), -\pi \leq x \leq \pi, \tag{24}$$

and zero elsewhere. In (24), $\eta$ is taken as a *real* variational parameter. Using (23) and (24) in (17), we quote the exact value and the best result obtained numerically:

$$\Delta(x^2)\Delta p = \tfrac{1}{2} \geq 0.395 \ (\eta = 1). \tag{25}$$

This is, admittedly, quite good a bound. As an interesting alternative, one might opt for (15) by choosing two separately normalized states that are parts of (24), *e.g.*,

$$\Psi_{N1} = \tfrac{1}{\sqrt{\pi}} \cos \tfrac{3}{2}x, -\pi \leq x \leq \pi,$$
$$\Psi_{N2} = \tfrac{1}{\sqrt{\pi}} \sin x, -\pi \leq x \leq \pi,$$

and zero otherwise. It is likely to offer a better estimate because of the greater flexibility. However, here we obtain the same estimate as was found in (25) using (24) [for a proof, see Appendix B]. The strength of (17), in spite of its simplicity, is obvious now.

Owing to symmetry between $x$ and $p$ in (22), the case of $A = p^2$ and $B = x$ for the above oscillator problem yields the same final result, *viz.*,

$$\Delta(p^2)\Delta x = \tfrac{1}{2} \geq 0.395 \ (\eta = 1). \tag{26}$$

Satisfaction of (12) and consequent trivialization of (2) in example 2 discussed above is quite general to all one-dimensional bound-state problems wherever the potential has *reflection* symmetry. In fact, the problem extends to all even powers of $x$ and odd powers of $p$, or the converse, provided the corresponding quantities exist. One may now appreciate the constructive role of (17) in all these situations.

### *3. Use of symmetry*

Example 2 also shows us a scheme of bypassing (17), thus requiring *no auxiliary state*. Let us note that the Cauchy-Schwarz inequality, displayed in (2), can be generally presented in the form

$$\left[\int_{x_1}^{x_2}(\Psi_A)^2 dx\right]\left[\int_{x_1}^{x_2}(\Psi_B)^2 dx\right] \geq \left[\int_{x_1}^{x_2}\Psi_A \Psi_B dx\right]^2$$

for two states $\Psi_A$ and $\Psi_B$ that are real functions (like here). Written in this form, one gains the added flexibility of tightening the inequality by choosing suitable pair of values for $(x_1, x_2)$. The only problem is to correlate the left side with observed spreads of observables $A$ and $B$. In order to achieve this end, and to simultaneously ensure a non-zero value for the right side, a simple escape route is to *limit the integration* within $(0, \infty)$. The approach halves each variance at the left, yields a finite answer for the right side, yet does not require the use of *any auxiliary state*. Thus, we do have really a nontrivial answer for problems like the ones discussed in example 2. For instance, in either of the two cases, result turns out to be

$$\Delta A \Delta B > \left| \int_0^\infty \phi_1 \phi_2 dx \right| = \tfrac{1}{2\sqrt{\pi}}. \qquad (27)$$

While the outcome is not as impressive as in (25) or (26), it is analytically tractable. The route will work in all other similar situations too, *e.g.*, those indicated below Eq. (26). What is more, this should count as an extra advantage of function space that is possibly lacking in the general vector space formalism.

### III. CONCLUSION

In fine, we observe the following. First, while the 'sum' forms (4) and (5) are correct, they suffer from the disadvantage that, unlike (2), they do not differentiate linearly dependent $\Psi_A$ and $\Psi_B$ from the linearly independent ones. Second, the nontrivial Case 2 yields from (4) or (5) almost a trivial result like (6) or (7), though we have later made profitable use of it. Third, the experimental justifications of the new inequalities with qubits/qutrits used actually linearly dependent $\Psi_A$ and $\Psi_B$, where the *full potential* of (2) is indeed apparent *almost* everywhere. Fourth, two simple routes are put forward in Sec. IIB3 and IIB4 to tackle the trouble with form (2) in Case 2; one of them does not involve any auxiliary state, and is noted to be useful elsewhere too. Fifth, and most important, we have found (17) in Sec. IIC1 to work quite satisfactorily in Case 3, requiring just *one* auxiliary state; it can be extended to include multiple incompatible observables as well. Sixth, *en route*, we have also simplified forms (4) and (5) through (14) and (16), of which the latter is found far more efficient. Seventh, in function space, we have seen in Sec. IIC3 how (2) can be forced to work using a convenient subspace, even when (12) holds, *without the aid of any auxiliary state*. This is also very different in spirit from (4) or (5). Finally, it may be truly challenging to have a different and 'stronger uncertainty relation' that is able to tackle (2) in Case 3, but neither invoking the function-space advantage in the way we did, nor importing any *auxiliary* state.

## APPENDIX A: SPECIALTY OF THE EXPERIMENTAL STATES

Let us start by defining the following normalized angular momentum states $\Phi$ and some of their properties of current concern:

$$\begin{aligned}
J^2 \Phi(j,\mu) &= j(j+1)\,\Phi(j,\mu), \\
J_z \Phi(j,\mu) &= \mu\,\Phi(j,\mu), \\
J_+ \Phi(j,\mu) &= \sqrt{(j-\mu)(j+\mu+1)}\,\Phi(j,\mu+1) \\
J_- \Phi(j,\mu) &= \sqrt{(j+\mu)(j-\mu+1)}\,\Phi(j,\mu-1)
\end{aligned} \quad (A1)$$

Expressing $J_x$ and $J_y$ in terms of $J_+$ and $J_-$, one then obtains

$$\begin{aligned}
\langle \Phi(j,\mu) | J_x | \Phi(j,\mu) \rangle &= 0, \\
\langle \Phi(j,\mu) | J_y | \Phi(j,\mu) \rangle &= 0.
\end{aligned} \quad (A2)$$

For some normalized state $\Psi$ written as a linear combination of two such $\Phi$ states, *viz.*,

$$\Psi = \frac{1}{\sqrt{|c_1|^2 + |c_2|^2}}[c_1 \Phi(j,\mu_1) + c_2 \Phi(j,\mu_2)], \quad (A3)$$

it is also easy to arrive at the following relation

$$\langle \Psi | J_x | \Psi \rangle = 0 = \langle \Psi | J_y | \Psi \rangle, \, |\mu_1 - \mu_2| \neq 1. \quad (A4)$$

A special case of (A3) in the form

$$\Psi = \frac{1}{\sqrt{|c_1|^2 + |c_2|^2}}[c_1 \Phi(j,\mu) + c_2 \Phi(j,-\mu)] \quad (A5)$$

is additionally a simultaneous eigenstate of $J^2$ and $J_z^2$, and hence of $(J_x^2 + J_y^2)$. We now discuss special features of the two choices for $\Psi$ given in (8) and (10).

**(i) Qutrits**

Consider first the specific linear combination that defines the state (8), *i.e.*,

$$\Psi = [\sin\varphi\,\Phi(1,1) + \cos\varphi\,\Phi(1,-1)]. \quad (A6)$$

By virtue of (A4), the state (A6) yields

$$\langle J_x \rangle = 0 = \langle J_y \rangle. \quad (A7)$$

Further, this state has also the form of (A5). So, it satisfies the eigenvalue equations

$$\begin{aligned}
J^2 \Psi &= 2\Psi, \\
J_z^2 \Psi &= \Psi.
\end{aligned} \quad (A8)$$

As a result, (A6) also obeys

$$(J_x^2 + J_y^2)\Psi = (J^2 - J_z^2)\Psi = \Psi. \tag{A9}$$

Therefore, in view of (A7) and (A9), one is led to the result

$$(\Delta J_x)^2 + (\Delta J_y)^2 = 1. \tag{A10}$$

Note that (A10) is obtained without calculating *any actual variance*, and this result is true for any $\varphi$. Indeed, if experiments on state (A6) merely ensure the average results (A7), and the eigenvalues that appear in (A8), one would automatically be led to (A10). Thus, the employment of (A6) in the context of experimentally demonstrating any *general* uncertainty relation like (4) becomes very special. An added specialty of the choice (A6) concerns the action of $J_x$ or $J_y$. One finds

$$\begin{aligned}J_x[\sin\varphi\,\Phi(1,1)+\cos\varphi\,\Phi(1,-1)]&=(1/\sqrt{2})(\cos\varphi+\sin\varphi)\Phi(1,0)\\ J_y[\sin\varphi\,\Phi(1,1)+\cos\varphi\,\Phi(1,-1)]&=(1/\sqrt{2}i)(\cos\varphi-\sin\varphi)\Phi(1,0)\end{aligned}. \tag{A11}$$

It shows that (i) $J_x\Psi$ and $J_y\Psi$ are generally linearly dependent, and (ii) $J_y\Psi = 0$ at $\varphi = \pi/4$ so that the linear dependence is lost at this point. It is also remarkable that this choice of $\varphi$ in (A6) yields the interesting relations

$$\begin{aligned}J_x\Psi &= \Phi(1,0),\ J_x^2\Psi = \Psi;\\ J_x\Phi(1,0) &= \Psi,\ J_x^2\Phi(1,0) = \Phi(1,0).\end{aligned} \tag{A12}$$

Such outcomes imply that $J_x$ acts as a *toggle operator* for the particular pair of states. This is rarely observed. Two specific linear combinations of the pair of degenerate eigenstates of $J_x^2$ shown in (A12) yield the eigenstates of $J_x$ with eigenvalues ±1. This explains how, in finite dimensions, it is possible that there exist common eigenstates of operators $X^2$ and $Y^2$, though $[X,Y] \neq 0$. Here, $J_x$ and $J_z$ act as these two noncommuting operators (for a simpler case, see below). Similar observations follow for the operator $J_y$ at $\varphi = 3\pi/4$ where $J_x\Psi = 0$. Let us note in passing that these outcomes are very special of $\Phi(1,\pm 1)$ states used in (A6).

**(ii) Qubits**

Qubits possess a special advantage primarily because of the peculiar relations among its operators like

$$[\sigma_x,\sigma_y] \neq 0;\ [\sigma_x^2,\sigma_y^2] = 0, \tag{A13}$$

the latter having its origin in $\sigma_x^2 = \sigma_y^2 = \sigma_z^2 = I$. Consider now the specific linear combination that defines the state (10), *i.e.*,

$$\Psi = [\sin\varphi\, \Phi(\tfrac{1}{2},\tfrac{1}{2}) + \cos\varphi\, \Phi(\tfrac{1}{2},-\tfrac{1}{2})]. \tag{A14}$$

This has the form of (A3), though (A4) is not valid and that's why we found $\langle\sigma_x\rangle \neq 0$. However, (A14) satisfies (A5). Hence, the state is a simultaneous eigenstate of $\sigma^2$ and $\sigma_z^2$. As a result, we obtain like (A9) the result

$$\left(\sigma_x^2 + \sigma_y^2\right)\Psi = 2\Psi. \tag{A15}$$

Coupled with the values of $\langle\sigma_x\rangle$ and $\langle\sigma_y\rangle$ given in (10), one obtains the final expression displayed there.

## APPENDIX B: THE AUXILIARY STATE AND CASE 3

Consider the construction of a complete orthonormal set $\{\theta_j\}$ from another such set by insisting that

$$\theta_1 = \Psi,\; \theta_2 = \overline{\Psi}_A,\; \theta_3 = \overline{\Psi}_B. \tag{B1}$$

In Case 3 [see Eqs. (1) and (12)], such choices are always permissible and other states can be accordingly built using suitable projection operators. A general auxiliary state, as used in the text, may then be written as

$$\Psi_N = \frac{1}{\sqrt{\sum_j r_j^2}} \sum_j c_j \theta_j ;\, |c_j| = r_j. \tag{B2}$$

Thus, the inequalities in (7) and in (13) – (16) are saturated. It also follows from (B2) that the following quantity in (17) is our prime concern:

$$\left|\langle\Psi_A | \Psi_N\rangle\right|\left|\langle\Psi_B | \Psi_N\rangle\right| = \frac{r_2 r_3}{\sum_j r_j^2} \Delta A \Delta B. \tag{B3}$$

The independent linear variational parameters $\{r_j\}$ at the right side of (B3) may now be varied to maximize the concerned measure, and this exercise leads to the results

$$r_j = 0,\, j \neq 2,3;\, r_2 = r_3. \tag{B4}$$

After optimization, we thus obtain from (B3)

$$\left[|\langle \Psi_A | \Psi_N \rangle||\langle \Psi_B | \Psi_N \rangle|\right]_{max} = \frac{1}{2}\Delta A \Delta B. \tag{B5}$$

Since this is the maximum possible value of the product at the left of (B3), the inequality in (17) follows. The same procedure would lead to the generalization for $n$ orthonormal states $\bar{\Psi}_A, \bar{\Psi}_B, \bar{\Psi}_C, \ldots$, as displayed below Eq. (17).

A related and interesting point is how (15) and (17) can yield the same final result. This was noted in Sec. IIC2 around Eq. (25). Indeed, when $\Psi_{N1}$ and $\Psi_{N2}$ in (15) are orthogonal and one expresses $\Psi_N$ in (17) as

$$\Psi_N = \frac{1}{\sqrt{r_1^2 + r_2^2}}(c_1 \Psi_{N1} + c_2 \Psi_{N2}), \; r_j = |c_j|, \tag{B6}$$

it turns out that an optimization as mentioned above leads again to $r_1 = r_2$. Consequently, one obtains the same estimate that has been found from (15). This means, a simple equal linear combination in (17), like the choice

$$\Psi_N = \frac{1}{\sqrt{2}}(\Psi_{N1} + \Psi_{N2}), \tag{B7}$$

serves the purpose best.

the average energy yields $k = (3\lambda)^{1/3}$ and, we note, this *optimal scaling* leads to the result $\langle T \rangle = 2 \langle V \rangle$. Thus, the virial theorem is *exactly* satisfied here. However, additional calculations also reveal that $\Delta T = (3\lambda)^{1/3}/\sqrt{2}, \Delta V = 2\Delta T/\sqrt{3}.$ These results show quite transparently that, not only $\Delta T \neq \Delta V$ is true at any $\lambda > 0$, but $|\Delta T - \Delta V|$ also increases as $\lambda$ rises, worsening gradually the quality of the function as a stationary state. This is how one can test the 'goodness' of approximate stationary states via our prescription.

[37] P. O. Lowdin, J. Mol. Spectrosc. **3**, 46 (1959).